\documentclass[aps, prb, nolongbibliography, reprint]{revtex4-2}
\newcommand{\thedoctitle}{Strictly Localized Mixed States}
\newif\ifhyperstyle\hyperstyletrue
\newif\ifextraconfig\extraconfigtrue
\newif\iffigfrompdf\figfrompdffalse
\usepackage{setup}

\begin{document}

\title{\thedoctitle}

\author{Sigurd S. Rustad}
\affiliation{Department of Physics, University of Oslo, NO-0316 Oslo, Norway}

\author{Jan Gulla}
\affiliation{Department of Physics, University of Oslo, NO-0316 Oslo, Norway}

\author{Johannes Skaar}
\email{johannes.skaar@fys.uio.no}
\affiliation{Department of Physics, University of Oslo, NO-0316 Oslo, Norway}

\date{\today}

\begin{abstract}
We study strictly localized mixed states in quantum field theory: states that are indistinguishable from vacuum outside some spacetime region. Such states are generated from the vacuum by so-called Licht maps. While the states associated with individual outcomes of local, projective measurements on the vacuum are not strictly localized, the corresponding non-selective states are strictly localized and even admit decompositions into strictly localized pure states. Remarkably, however, there are strictly localized mixed states that cannot be written as mixtures of pure states strictly localized to the same region. Such states can arise from unsharp local measurements. Finally, we extend Knight's theorem to mixed states, showing that strictly localized bosonic mixed states contain terms with arbitrarily high particle numbers.
\end{abstract}

\maketitle

\section{Introduction}
Local measurements have been widely studied in quantum field theory \cite{haag1964, haag1996, de_bievre2006, brunetti2015, witten2018, falcone2024, gulla:inprep:a}. In the algebraic formulation, locality is encoded by assigning to each spacetime region an algebra of local operators, interpreted as describing physical observables and operations available there. Building on this, two quantum states are locally equivalent in a region if they give the same expectation value for every observable in the corresponding local algebra.

Strict localization is defined by comparing expectation values with those of the vacuum. Specifically, a state is strictly localized to a given region if it gives the same expectation values as the vacuum for every local observable outside that region \cite{knight1961}. In other words, no outside measurement can then distinguish the state from the vacuum. Strict localization therefore provides an operational criterion for a state change to be localized consistently with causality, formulated in terms of local observables rather than a tensor-product decomposition of the Hilbert space. 

Because the vacuum exhibits nontrivial correlations between spacelike-separated regions \cite{reeh1961, witten2018}, the distinction between a global state change and its local detectability can be subtle. To gain some intuition for how locality can work in quantum field theory, it is useful to consider an analogy from non-relativistic quantum mechanics \cite{redhead1995, wallace2006}. Consider a two-qubit system, with one qubit held by Alice and the other by Bob. We take the analog of the vacuum to be the maximally entangled Bell state
\begin{equation}
    \ket{\Omega} = \frac{\ket{00} + \ket{11}}{\sqrt{2}}.
\end{equation}
An analog of strict localization to Alice's system is obtained by requiring that a state have the same reduced state on Bob's system as the vacuum. With this definition, for example, states of the form $\rho_A\otimes I_B/2$ are strictly localized to Alice's system. However, not all strictly localized states are of this form. For example, the vacuum state $\ket{\Omega}\bra{\Omega}$ itself cannot be written as a tensor product state across Alice's and Bob's systems. Hence, states localized to Alice's system should be regarded as a subset of global states of the joint system rather than as reduced states of Alice's subsystem alone.

We next consider what happens when Alice performs a local measurement on her system. A measurement in the computational basis might, for example, yield the outcome ``1'', in which case the selective post-measurement state is $\ket{\Omega}\mapsto\ket{11}$. This state is not strictly localized to Alice's system, because Alice knows with certainty that Bob measuring in the same basis will also obtain the outcome ``1''. Even so, no information is transmitted through this state collapse, because the corresponding non-selective state after Alice's measurement is the mixed state
\begin{equation}
    \rho = \frac{1}{2}\ket{00}\bra{00} + \frac{1}{2}\ket{11}\bra{11},
\end{equation}
which gives Bob the same reduced state as if Alice had done nothing. That is,
\begin{equation}\label{eq_intro_sl}
    \tr_A(\rho) = \tr_A(\ket{\Omega}\bra{\Omega}) = \frac{I_B}{2},
\end{equation}
where $\tr_A$ denotes the partial trace over Alice's system, and $I_B$ is the identity operator on Bob's system. Thus, although the global state changed, this change was not locally observable by Bob.

We see that the strictly localized state $\rho$ can be written as a mixture of pure states that are not strictly localized. One may ask whether other ensemble decompositions of $\rho$ exist, in particular decompositions into pure states that are strictly localized. In this case, such a decomposition does exist, as
\begin{equation}
    \rho = \frac{1}{2} \ket{\Phi_+}\bra{\Phi_+} + \frac{1}{2} \ket{\Phi_-}\bra{\Phi_-},
\end{equation}
for the Bell states $\ket{\Phi_{\pm}} = \lp(\ket{00} \pm \ket{11})/\sqrt{2}$. 

The same ideas carry over to quantum field theory: localization is characterized by how the state looks outside the region, and local operations may genuinely change the global state \cite{reeh1961}. One difference is that, whereas measurement statistics on Bob's side in the qubit model are encoded by his reduced state, no analogous reduced density operator is generally available in quantum field theory. This is because the local algebras associated with a region and its complement do not define tensor factors of the Hilbert space. One instead requires agreement with the vacuum on all local observables outside the region.

The relation between strictly localized pure states and states produced from the vacuum by local on-demand sources has been studied, e.g., in Refs.~\cite{gulla2021, gulla:inprep:a}. Here, on demand means that a fixed local preparation procedure produces the desired pure state with unit probability. Mixed states, on the other hand, arise naturally from measurements, when measurement outcomes are ignored or an auxiliary probe system is discarded.

The paper is organized as follows. \Cref{sec_2} introduces strict localization, first for pure states and then for mixed states. In \cref{sec_3}, we show that local, projective measurements on the vacuum necessarily produce selective states that are not strictly localized. Nevertheless, the non-selective mixed state is strictly localized, and can in fact always be expressed as a mixture of strictly localized pure states. This naturally raises the question: are there strictly localized mixed states that cannot be written as mixtures of strictly localized pure states for a given region? In the first part of \cref{sec_4}, we answer this question affirmatively. In the second part, we show that some of these states cannot even be approximated by mixtures of strictly localized pure states. In \cref{sec_5}, we discuss how such states can be interpreted in terms of more general, unsharp measurement schemes. Finally, \cref{sec_6} contains our concluding remarks.

\section{Strictly Localized States\label{sec_2}}

Given a region $\mathcal{O}$ in Minkowski spacetime, we have a corresponding local algebra $\mathcal{A}(\mathcal{O})$ \cite{haag1964, haag1996, de_bievre2006, halvorson2007, brunetti2015, witten2018, falcone2024, gulla:inprep:a}, which can be thought of as an algebra containing observables and operators local to $\mathcal{O}$. As a concrete example, consider a free hermitian scalar field $\phi(x)$ and a complex smearing function $f(x)$ with support in $\mathcal{O}$. We define smeared fields as
\begin{equation}
    \phi_f = \int \dd[4]x \, f(x)\phi(x).
\end{equation}
The local algebra can then be taken to consist of sums and products of such smeared fields with different smearing functions. However, since smeared fields are unbounded and have unwieldy domains, one usually considers bounded versions of them instead. These can be obtained by applying bounded functions, such as $e^{i\phi_f}$ for real $f(x)$; see \cite{fewster2019} for further examples.

Given a Hilbert space $H$, Knight originally defined strict localization for pure states $\ket{\psi}\in H$ \cite{knight1961}. Specifically, $\ket{\psi}$ is strictly localized to $\mathcal{O}$ if
\begin{equation}
    \bra{\psi}A\ket{\psi} = \bra{\Omega}A\ket{\Omega}, \quad \forall A\in\mathcal{A}(\mathcal{O}^C),
\end{equation}
where $\ket{\Omega}$ is the vacuum and $\mathcal{O}^C$ denotes the complement of $\mathcal{O}$ in Minkowski spacetime. Equivalently, it suffices to impose this condition on self-adjoint operators $A\in\mathcal{A}(\mathcal{O}^C)$, which represent local observables \cite{gulla:inprep:a}. In other words, the states strictly localized to $\mathcal{O}$ are those that cannot be distinguished from the vacuum by local measurements outside $\mathcal{O}$. Note that some localization regions have only the vacuum state localized to them. In particular, Knight has shown that any state other than the vacuum cannot have a localization region that is bounded in time, due to propagation \cite{knight1961}.

Licht characterized how to generate strictly localized states from the vacuum \cite{licht1963}. Define $\mathcal{A}(\mathcal{O})'$ as the commutant of $\mathcal{A}(\mathcal{O})$, i.e., the set of operators that commute with every element in $\mathcal{A}(\mathcal{O})$. Then, for each state $\ket{\psi}$ strictly localized to $\mathcal{O}$, there exists an isometric operator $W\in \mathcal{A}(\mathcal{O}^C)'$ such that $W\ket{\Omega}=\ket{\psi}$. We call these isometries Licht operators associated with $\mathcal{O}$, and they are the only operators in $\mathcal{A}(\mathcal{O}^C)'$ that generate strictly localized states from the vacuum. Indeed, let $W\in\mathcal{A}(\mathcal{O}^C)'$ and suppose that $W\ket{\Omega}$ is strictly localized to $\mathcal{O}$. Then, by strict localization,
\begin{equation}
    \bra{\Omega}A^\dagger(W^\dagger W-I)B\ket{\Omega}=0, \quad \forall A,B\in\mathcal{A}(\mathcal{O}^C).
\end{equation}
By the Reeh--Schlieder property \cite{reeh1961}, the vanishing of all these matrix elements implies that $W^\dagger W =I$. Alternatively, this means that if $W\in\mathcal{A}(\mathcal{O}^C)'$ and $W^\dagger W \neq I$, then $W\ket{\Omega}$ is not strictly localized to $\mathcal{O}$.

Later, these results were generalized to mixed states as well \cite{licht1966}. However, since the work was not formulated in terms of (quantum) operations \cite{hellwig1969}, we also refer to Ref.~\cite{hellwig1970}. For a state represented by a density operator $\rho$, we say that $\rho$ is strictly localized to $\mathcal{O}$ if
\begin{equation}\label{eq_sl_definition}
    \tr(A\rho) = \tr(A\Omega), \quad \forall A\in\mathcal{A}(\mathcal{O}^C),
\end{equation} 
where $\Omega = \ket{\Omega}\bra{\Omega}$.

To extend Licht operators to mixed states, it is natural to use operations. Recall that an operation is a completely positive, trace non-increasing map. Every such map $\mathcal{W}$ admits a Kraus representation \cite{kraus1971}
\begin{equation}\label{eq_kraus_rep}
    \mathcal{W}(\rho) = \sum_k W_k \rho W_k^\dagger,
\end{equation}
where the Kraus operators $W_k$ satisfy
\begin{equation}
    \sum_k W_k^\dagger W_k \leq I.
\end{equation}
The operation is non-selective, or trace preserving, whenever equality holds. We say that $\mathcal{W}$ is an operation local to $\mathcal{O}$ if its Kraus operators may be chosen such that $W_k \in \mathcal{A}(\mathcal{O})$ \cite{haag1964, kraus1971}; see also Refs.~\cite{okamura2015, okamura2021, fewster2020, fewster2024, fewster2025} for recent developments on local operations. 

\begin{figure}[tbp]
\centering
\resizebox{0.95\linewidth}{!}{%
\begin{tikzpicture}[>=Latex]
    \def\a{1.0}   
    \def\T{2.3}   
    \def\X{4.6}   

    \draw[->] (-\X,0) -- (\X,0);
    \draw[->] (0,-\T) -- (0,\T);

    \fill[gray!15]
        (-\X,\T) -- (-\a-\T,\T) -- (-\a,0) -- (-\a-\T,-\T) -- (-\X,-\T) -- cycle;
    \fill[gray!15]
        (\X,\T) -- (\a+\T,\T) -- (\a,0) -- (\a+\T,-\T) -- (\X,-\T) -- cycle;

    \fill[gray!22]
        (-\a-\T,\T) -- (\a+\T,\T) -- (\a,0) -- (\a+\T,-\T)
        -- (-\a-\T,-\T) -- (-\a,0) -- cycle;

    \draw[thick, gray!65!black] (-\a,0) -- (-\a-\T,\T);
    \draw[thick, gray!65!black] (\a,0) -- (\a+\T,\T);
    \draw[thick, gray!65!black] (-\a,0) -- (-\a-\T,-\T);
    \draw[thick, gray!65!black] (\a,0) -- (\a+\T,-\T);

    \filldraw[fill=blue!28, draw=blue!55!black, thick]
        (-\a,0)
        .. controls ({-0.72*\a},0.40) and ({0.72*\a},0.40) .. (\a,0)
        .. controls ({0.72*\a},-0.40) and ({-0.72*\a},-0.40) .. (-\a,0)
        -- cycle;

    \node at (0,0) {$\mathcal{D}$};
    \node[align=center] at (0,1.75) {$J(\mathcal{D})$};
    \node[align=center] at (0,-1.75) {$J(\mathcal{D})$};
    \node[align=center] at (-3,0) {$\mathcal{D}'$};
    \node[align=center] at (3,0) {$\mathcal{D}'$};
\end{tikzpicture}
}
\caption{Sketch in $1+1$ dimensions, with time increasing upwards, showing a bounded spacetime region $\mathcal{D}$, its causal complement $\mathcal{D}'$, and its causal future and past, denoted by $J(\mathcal{D})$.\label{fig_space_regions}}
\end{figure}
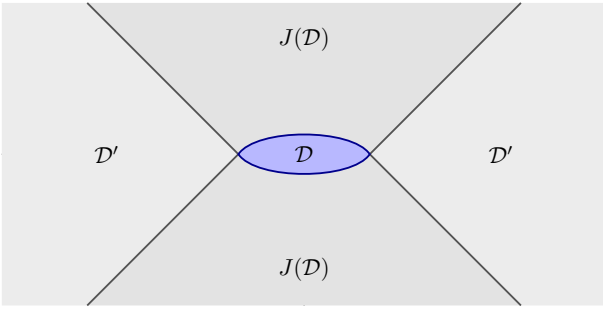

An immediate class of strictly localized mixed states is obtained by applying a non-selective local operation to the vacuum. Indeed, let $\mathcal{W}$ be a non-selective operation local to some bounded spacetime region $\mathcal{D}$. Then
\begin{equation}
    \rho = \mathcal{W}(\Omega)
\end{equation}
is strictly localized to $J(\mathcal{D})$, i.e., the set of points causally connected to $\mathcal{D}$. This follows because $\mathcal{W}$ admits the Kraus representation \eqref{eq_kraus_rep}, whose Kraus operators commute with every operator in $\mathcal{A}(\mathcal{D}')$, where $\mathcal{D}'$ is the set of all points spacelike separated from $\mathcal{D}$. Hence, $\rho$ satisfies \eqref{eq_sl_definition} with localization region $\lp(\mathcal{D}')^C = J(\mathcal{D})$. See \cref{fig_space_regions} for a sketch of the different spacetime regions.

Licht showed that every strictly localized state can be generated from the vacuum by a suitable non-selective operation. That is, if $\rho$ is strictly localized to $\mathcal{O}$, then there exists a non-selective operation $\mathcal{W}$ with Kraus operators $W_k \in \mathcal{A}(\mathcal{O}^C)'$ such that
\begin{equation}
    \rho =  \mathcal{W}(\Omega) = \sum_k W_k \Omega W_k^\dagger, \quad \sum_k W_k^\dagger W_k = I.
\end{equation}
We call such operations Licht maps associated with $\mathcal{O}$. Strictly speaking, Licht maps are not necessarily local operations in the sense defined above, since taking the commutant of a local algebra does not necessarily yield another local algebra. Certainly, $\mathcal{A}(\mathcal{O}')$ is a subset of $\mathcal{A}(\mathcal{O})'$; however, equality, known as Haag duality \cite[Chap. III.4]{haag1996}, does not hold in general.

Knight demonstrated that every bosonic, non-vacuum, strictly localized pure state must be a superposition of particle states with arbitrarily large particle numbers; see also \cite{de_bievre2006, gulla:inprep:a}. This does not mean that the expected particle number is infinite or even large. For example, coherent states can be strictly localized; they are superpositions of number states from zero to infinity, but their expected particle number can be any number. Also, there exist strictly localized states close to single particles \cite{gulla2021, ryen2022, gulla2023}.

In \cref{app_1} we demonstrate that Knight's result generalizes to strictly localized mixed states. That is, let $\rho$ be a strictly localized bosonic state other than the vacuum state. We may write it in the following form: 
\begin{equation}
    \rho = \sum_{n,m}\rho_{nm}, \quad \rho_{nm} = P_n\rho P_m,
\end{equation}
where $P_n$ is the projector onto the $n$-particle subspace. If there is an integer $N$ such that $\rho_{nm} = 0$ for $n>N$ or $m>N$, then $\rho$ cannot be strictly localized.

Since Licht maps can be represented by Kraus operators that commute with all observables outside the localization region, it is natural to ask whether this commutation property is intrinsic to the operation or whether it depends on the chosen Kraus representation. In \cref{app_2}, we study this question through the lens of fixed points and show that every Kraus representation of a Licht map $\mathcal{W}$ associated with $\mathcal{O}$ consists of operators $W_k\in\mathcal{A}(\mathcal{O}^C)'$. Thus, Licht maps may equivalently be viewed as non-selective operations that leave all observables outside the localization region unchanged in the fixed-point sense.

In the special case of pure states, we recover the original results of Knight and Licht. Thus, the mixed-state definition is the natural generalization of strict localization. The following sections investigate to what extent this generalization gives us a genuinely larger set of states.

\section{Local Projective Measurements and Decoherence\label{sec_3}}

In this section, we study the states produced by local measurements. We will show that the selective state after a local, projective measurement on the vacuum is not strictly localized. By contrast, the corresponding non-selective mixed state is strictly localized and, more surprisingly, can always be written as a mixture of strictly localized pure states.

Consider a bounded spacetime region $\mathcal{D}$ and a local, self-adjoint observable $A\in\mathcal{A}(\mathcal{D})$. By the spectral theorem, $A$ can be decomposed as
\begin{equation}
    A = \int_{\sigma(A)} \lambda \, dP(\lambda),
\end{equation}
where $P(\lambda)$ is a projection-valued measure, and $\sigma(A)$ is the spectrum of $A$. Operators with a discrete spectrum can be written
\begin{equation}\label{eq_A_disc}
    A_{\mathrm{disc}} = \sum_\lambda \lambda P_\lambda,
\end{equation}
with projectors $P_\lambda$, whereas operators with a continuous spectrum can be written
\begin{equation}\label{eq_A_cont}
    A_{\mathrm{cont}} = \int d\lambda \, \lambda P(\lambda),
\end{equation}
with a projector density $P(\lambda)$.

Initially, we consider $A$ to have a finite, discrete spectrum with at least two eigenvalues:
\begin{equation}
    A = \sum_{k=0}^{N-1} k P_k, \quad N \geq 2.
\end{equation}
This corresponds to a nontrivial projective measurement \cite{ozawa1984, ozawa1985}. Relabeling the eigenvalues changes the operator mathematically, but it represents the same physical observable as long as we do not introduce any new degeneracy. We will briefly revisit the infinite discrete and continuous scenarios at the end of this section. 

The spectral projectors satisfy $P_k = P_k^2 = P_k^\dagger$ and $P_kP_l =\delta_{kl}P_k$. If the initial state is the vacuum state $\Omega = \ket{\Omega}\bra{\Omega}$, then for a measured value $k$, the selective post-measurement state is (Lüders rule):
\begin{equation}\label{eq_psi_collapse}
    \ket{\psi} = \frac{P_k\ket{\Omega}}{\norm{P_k\ket{\Omega}}},    
\end{equation}
with a corresponding probability $\norm{P_k\ket{\Omega}}^2$ (Born rule). Note that by spectral functional calculus, we can obtain $P_k$ from $A$ by mapping the eigenvalue $k$ to one and all other eigenvalues to zero. Hence, $A\in\mathcal{A}(\mathcal{D})$ implies that $P_k\in\mathcal{A}(\mathcal{D})$. Since local operators cannot annihilate the vacuum \cite{reeh1961} (Reeh--Schlieder property), the probability $\norm{P_k\ket{\Omega}}^2$ is always greater than zero. This means that the state in \eqref{eq_psi_collapse} cannot be strictly localized. Indeed, the only operators in $\mathcal{A}(\mathcal{D}')'$ that generate strictly localized states from the vacuum are Licht operators, whereas $P_k/\norm{P_k\ket{\Omega}}$ is not an isometry and hence not a Licht operator. This means that for an observer who knows the measurement result of the local, projective measurement, the resulting state is not strictly localized. Note that this argument holds even if we take $N$ to be infinite.

On the other hand, the post-measurement state viewed from a spacelike separated observer must be strictly localized. Indeed, since the result is unknown for a spacelike separated observer, the state viewed from their perspective is given by the corresponding non-selective post-measurement state
\begin{equation}\label{eq_rho_finite_discrete}
    \rho = \sum_{k=0}^{N-1} P_k\Omega P_k.
\end{equation}
This is a local operation since each Kraus operator $P_k$ belongs to $\mathcal{A}(\mathcal{D})$ and $\sum_k P_k = I$. Hence, $\rho$ is strictly localized to $J(\mathcal{D})$, even though none of the pure states in the mixture are. 

Nevertheless, $\rho$ does admit a representation in terms of strictly localized pure states. Consider the $N\times N$ discrete Fourier transform matrix
\begin{equation}
    U = \frac{1}{\sqrt{N}}\begin{pmatrix}
1       & 1           & 1           & \cdots & 1 \\
1       & u           & u^{2}       & \cdots & u^{N-1} \\
1       & u^{2}       & u^{4}       & \cdots & u^{2(N-1)} \\
\vdots  & \vdots      & \vdots      & \ddots & \vdots \\
1       & u^{N-1}     & u^{2(N-1)}  & \cdots & u^{(N-1)(N-1)}
\end{pmatrix},
\end{equation}
where
\begin{equation}
    u = e^{-2\pi i/N}.
\end{equation}
This matrix is unitary. Define $W_l$ by
\begin{equation}
    W_l = \sqrt{N}\sum_{k=0}^{N-1} U_{lk}P_k = e^{-2\pi i l A / N}.
\end{equation}
It follows that each $W_l$ is a Licht operator associated with $J(\mathcal{D})$, and it therefore generates a state strictly localized to $J(\mathcal{D})$ from the vacuum. Also, since $U$ is unitary, we find that \eqref{eq_rho_finite_discrete} may be written in the following form:
\begin{equation}\label{eq_sl_rep}
    \rho = \frac{1}{N} \sum_{l = 0}^{N-1} W_l\Omega W^\dagger_l.
\end{equation}
In other words, $\rho$ may be expressed as a mixture of strictly localized pure states. 

We see that a large set of physically relevant strictly localized mixed states reduces to mixtures of strictly localized pure states. One might wonder if we can obtain more than just mixtures of strictly localized pure states. This is addressed in \cref{sec_4,sec_5}. The remainder of this section is used to discuss the interpretation of \eqref{eq_sl_rep} and generalizations to observables with other spectra. 

At first sight, the equivalence between \eqref{eq_rho_finite_discrete} and \eqref{eq_sl_rep} may seem surprising. However, it has a simple interpretation in terms of decoherence. Moreover, this interpretation provides insight into other types of spectra. To make this concrete, decompose the vacuum state in the following way:
\begin{equation}\label{eq_Omega_decomp}
    \Omega = \sum_{k,l} \Omega_{kl}, \quad \Omega_{kl}=P_k\Omega P_l.
\end{equation}
The off-diagonal blocks, $\Omega_{kl}$ with $k\neq l$, represent coherence between possible measurement outcomes of $A$ and are always nonzero since $P_k$ cannot annihilate the vacuum. In this context, decoherence means that $\Omega_{kl}\mapsto 0$ for $k\neq l$. The non-selective state in \eqref{eq_rho_finite_discrete} produces precisely this decoherence. We now show that \eqref{eq_sl_rep} gives an equivalent description in terms of phase averaging.

Note that by inserting \eqref{eq_Omega_decomp} into \eqref{eq_sl_rep}, each block transforms as:
\begin{equation}
    e^{-itA}\Omega_{kl}e^{itA} = e^{-it(k-l)}\Omega_{kl}.
\end{equation}
The off-diagonal blocks acquire phases, but the diagonal blocks do not. An averaging over the phases, therefore, removes the off-diagonal blocks while retaining the diagonal ones. Concretely, we obtain the following decomposition into strictly localized pure states:
\begin{equation}\label{eq_haar_twirl}
    \rho = \int_{-\pi}^\pi \frac{dt}{2\pi} \, e^{-itA} \Omega e^{itA}.
\end{equation}
For a finite integer-valued spectrum, this integral agrees exactly with \eqref{eq_sl_rep}. However, the expression remains valid for an infinite integer-valued spectrum as well. For more general discrete spectra, we may relabel the distinct eigenvalues to integers without changing the spectral projectors and recover \eqref{eq_haar_twirl}. Hence, the decomposition into strictly localized states remains valid for any discrete observable. Formally, the phase average in \eqref{eq_haar_twirl} is known as a group twirl \cite{Bartlett2007} or Haar average \cite{emerson2005}. 

For an observable with a continuous spectrum, it is impossible to relabel the spectrum to discrete values without introducing degeneracy and thereby changing the spectral projectors. However, we will see that several ideas generalize. As an example, consider a smeared field $\phi_f$, where $f(x)$ is a real smearing function with support in a bounded region $\mathcal{D}$. By spectral functional calculus, we can consider the unitary operator
\begin{equation}\label{eq_exp_phi_f}
    e^{-it\phi_f} = \int_{-\infty}^\infty d\lambda \, e^{-it\lambda} P(\lambda),
\end{equation}
where $t$ is a real parameter, and $P(\lambda)$ is a projector density. The continuous version of \eqref{eq_Omega_decomp} may be written as
\begin{equation}\label{eq_Omega_decomp_cont}
    \Omega = \int_{-\infty}^\infty d\lambda\, d\lambda' \, \Omega_{\lambda\lambda'}, \quad \Omega_{\lambda\lambda'} = P(\lambda)\Omega P(\lambda').
\end{equation}
Furthermore, a natural generalization of the phase average in \eqref{eq_haar_twirl} is
\begin{equation}\label{eq_weighted_twirl}
    \rho = \int_{-\infty}^{\infty}dt\, p(t) e^{-it\phi_f} \Omega e^{it\phi_f},
\end{equation}
where the constant factor $1/2\pi$ has been replaced by a probability density $p(t)$, which we will see provides varying degrees of decoherence. This state is a mixture of strictly localized pure states, each localized to $J(\mathcal{D})$. By inserting \eqref{eq_Omega_decomp_cont}, we obtain
\begin{equation}\label{eq_cont_haar_twirl}
    \rho = \int_{-\infty}^\infty d\lambda \, d\lambda' \, \hat{p}(\lambda - \lambda')\Omega_{\lambda\lambda'},
\end{equation}
where
\begin{equation}
    \hat p(\lambda-\lambda') = \int_{-\infty}^{\infty}dt\,p(t)e^{-it(\lambda-\lambda')}
\end{equation}
is the characteristic function of $p(t)$. Thus, the diagonal blocks are unchanged since $\hat{p}(0) = 1$, whereas the off-diagonal blocks are generally suppressed, giving decoherence.

Equation \eqref{eq_cont_haar_twirl} also shows why we cannot have complete decoherence as in the discrete case. Demanding $\hat{p}(\lambda - \lambda') = \delta(\lambda - \lambda')$ is incompatible with $p(t)$ being a probability distribution. This illustrates the difficulty of extending Lüders' rule directly to observables with continuous spectra. The subtleties of state-update rules for observables with continuous spectra are well known \cite{davies1976, Busch2009, holevo2011, ozawa2012}, and the standard framework for treating them is provided by Davies--Lewis instruments \cite{davies1970}. Although the general framework is technical, this simple example recovers the non-selective channel of a familiar Davies--Lewis instrument. Concrete examples appear in the Fewster--Verch framework \cite{fewster2020, mandrysch2025} and in position decoherence \cite{sajnok2026}.

\section{The Space of Strictly Localized States\label{sec_4}}

In the previous section, we showed that a large class of strictly localized mixed states can be written in terms of strictly localized pure states. This might suggest that every strictly localized mixed state can be obtained in this way. In this section, we show that this is not the case. In \cref{sec_4_1}, we construct an explicit example of a strictly localized mixed state that cannot be expressed as a mixture of pure states strictly localized to the same region. In \cref{sec_4_2}, we strengthen this result by showing that the same state has a positive trace distance from any such mixture.

\subsection{A Strictly Localized Mixture of Non-Localized Pure States\label{sec_4_1}}

To construct our example, let $\mathcal{D}$ be a bounded spacetime region and consider four nonzero, mutually orthogonal projectors $P_{k}\in \mathcal{A}(\mathcal{D})$ satisfying $\sum_k P_k =I$. Define the following two Kraus operators:
\begin{equation}\label{eq_example_W}
    W_1 = \sum_k \alpha_k P_k, \ \ W_2 = \sum_k \beta_k P_k,
\end{equation}
where $\alpha_k$ and $\beta_k$ are coefficients so that $|\alpha_k|^2 + |\beta_k|^2 = 1$ for each $k=1,\ldots,4$. Then,
\begin{equation}
    W_1^\dagger W_1 + W_2^\dagger W_2 = \sum_k P_k = I.
\end{equation}
Thus, $W_1$ and $W_2$ define a Licht map associated with $J(\mathcal{D})$, and the state
\begin{equation}\label{eq_local_state}
    \rho = W_1\Omega W_1^\dagger + W_2\Omega W_2^\dagger
\end{equation}
is strictly localized to $J(\mathcal{D})$. In \cref{sec_5}, we discuss the physical interpretation of this operation.

As seen in \cref{sec_3}, the same density operator can admit many different decompositions into pure states. We want to check if $\rho$ admits a decomposition into strictly localized pure states, localized to the same region. Suppose, then, that
\begin{equation}\label{rho_psi_l}
    \rho = \sum_l p_l \ket{\psi_l}\bra{\psi_l}
\end{equation}
is an arbitrary decomposition of this form, where $p_l>0$ are probabilities. From \eqref{eq_local_state}, the range of $\rho$ is contained in the subspace spanned by $W_1\ket{\Omega}$ and $W_2\ket{\Omega}$. At the same time, a decomposition of the form \eqref{rho_psi_l} where any $\ket{\psi_l}$ has a component outside this subspace would give $\rho$ nonzero support there, since positivity prevents such contributions from canceling. This remains true for any ensemble decomposition of $\rho$, even if it is not a spectral decomposition and the states are not mutually orthogonal. Hence, every $\ket{\psi_l}$ must be of the form
\begin{equation}
    \ket{\psi_l} = a_lW_1\ket{\Omega}+b_lW_2\ket{\Omega} = \widehat{W}_l\ket{\Omega},
\end{equation}
for some coefficients $a_l$ and $b_l$, and where $\widehat{W}_l=a_lW_1+b_lW_2$. Equivalently, this conclusion can be viewed as an instance of the Schrödinger--Hughston--Jozsa--Wootters theorem \cite{schrodinger1936, hughston1993} and its infinite-dimensional generalizations \cite{kirkpatrick2006, hadjisavvas1981, halvorson2004, wang2014}; see, for example, \cite[Lemma 2.5]{wang2014}.

Note that each $\widehat{W}_l$ is in $\mathcal{A}(\mathcal{D})$ because $W_{1}$ and $W_{2}$ are. Since Licht operators are the only operators in $\mathcal{A}(\mathcal{D}')'$ that generate strictly localized states from the vacuum, $\ket{\psi_l}$ can be strictly localized only if $\rule{0pt}{11pt}\smash{{\widehat{W}_l\null}^{\!\dagger} \widehat{W}_l = I}$. Writing out $\rule{0pt}{11pt}\smash{{\widehat{W}_l\null}^\dagger \widehat{W}_l}$ and using \eqref{eq_example_W}, we obtain
\begin{equation}
    {\widehat{W}_l\null}^\dagger \widehat{W}_l = \sum_k \abs{a_l \alpha_k + b_l \beta_k}^2 P_k.
\end{equation}
Since the $P_k$ are mutually orthogonal, we find that $\rule{0pt}{11pt}\smash{{\widehat{W}_l\null}^\dagger \widehat{W}_l = I}$ is equivalent to
\begin{equation}\label{eq_strict_loc_condition}
    \abs{a_l \alpha_k + b_l \beta_k} = 1, \quad \forall k.
\end{equation}
For suitable $\alpha_k$ and $\beta_k$, this is impossible. For example, the following choices lead to a contradiction:
\begin{subequations}
\begin{align}
    (\alpha_1, \beta_1) = (1, 0) &\Rightarrow \abs{a_l}^2 = 1, \\
    (\alpha_2, \beta_2) = (0, 1) &\Rightarrow \abs{b_l}^2 = 1, \\
    (\alpha_3, \beta_3) = \frac{1}{\sqrt{2}}(1, 1) &\Rightarrow \Re{a_l^* b_l} = 0, \\
    (\alpha_4, \beta_4) = \frac{1}{\sqrt{2}}(1, i) &\Rightarrow \Im{a_l^* b_l} = 0.
\end{align}
\end{subequations}
The first two conditions imply $|a_l^*b_l|=1$, whereas the last two imply $a_l^*b_l=0$. Hence, we have constructed a simple example of a mixed state strictly localized to $J(\mathcal{D})$ that cannot be expressed as a discrete mixture of pure states strictly localized to $J(\mathcal{D})$. The following subsection will show a stronger result.

\subsection{Convex Hull Is Not Dense\label{sec_4_2}}

In the previous subsection, we constructed a strictly localized state that cannot be expressed as a mixture of pure states strictly localized in the same region. However, this alone does not exclude the possibility that it can be approximated arbitrarily well in trace distance by such mixtures. In this subsection, we rule out this possibility by showing that the constructed state remains at a strictly positive trace distance from every such mixture. We then discuss the physical interpretation of this separation.

Let $\mathcal{D}$ again be a bounded spacetime region. Let $\mathcal{R}$ denote the set of all (mixed) states strictly localized to $J(\mathcal{D})$, and let $\Psi\subseteq\mathcal{R}$ denote the subset of pure states. Take $\rho\in\mathcal{R}$ to be as in the previous subsection. Define the convex hull of $\Psi$ as all possible finite mixtures of its elements:
\begin{equation}
    \operatorname{co}(\Psi) = 
    \left\{
    \sum_{k=1}^{N} p_k\psi_k \,\middle|\,\psi_k\in\Psi,\; p_k\geq 0,\;\sum_{k=1}^{N}p_k=1
    \right\}.
    \end{equation}
By construction, $\Psi \subseteq \operatorname{co}(\Psi) \subseteq \mathcal{R}$. In the previous subsection, we showed that $\rho \notin \operatorname{co}(\Psi)$, so $\operatorname{co}(\Psi) \neq \mathcal{R}$. If $\operatorname{co}(\Psi)$ is closed in trace norm, then this would already prove that $\rho$ cannot be approximated by such mixtures. However, there is no guarantee that $\operatorname{co}(\Psi)$ is closed. Its closure may contain limits, including countable or uncountable mixtures of the kind encountered in \cref{sec_3}. Hence, we use $\overline{\operatorname{co}(\Psi)}$ to denote its closure. We will show that even this potentially larger set cannot approach $\rho$ in trace distance.

Let $W_1$ and $W_2$ be the Kraus operators used to construct $\rho$ in the previous subsection. The key observation is that the subspace generated by $W_1\ket{\Omega}$ and $W_2\ket{\Omega}$ is separated from all strictly localized pure states. Define $S= \operatorname{span}\{W_1\ket{\Omega}, W_2\ket{\Omega}\}$. We know that $\ket{\psi} \not\in S$ holds for all $\ket{\psi}\bra{\psi}\in \Psi$. Let $P_S$ be the orthogonal projector onto $S$. Then, there exists a constant $c<1$ such that
\begin{equation}\label{eq_positive_gap}
    \tr(P_S \psi)  \leq c,
\end{equation}
for all $\psi\in\Psi$. To see this, assume for contradiction that no such $c$ exists. Then there would be a sequence $\psi_k = \ket{\psi_k}\bra{\psi_k}\in\Psi$ such that 
\begin{equation}\label{eq_psi_k_sec}
    \tr(P_S \psi_k) = \norm{P_S \ket{\psi_k}}^2 \rightarrow 1.
\end{equation}
However, this implies that
\begin{equation}
    \norm{(I-P_S)\ket{\psi_k}}^2 = 1-\norm{P_S\ket{\psi_k}}^2 \rightarrow 0.
\end{equation}
Since $S$ is finite-dimensional and $P_S\ket{\psi_k}$ in \eqref{eq_psi_k_sec} is a norm-bounded sequence in $S$, we may pass to a subsequence such that
\begin{equation}
    P_S\ket{\psi_k}\rightarrow\ket{\psi},
\end{equation}
for some $\ket{\psi}\in S$.  It follows that, along this subsequence,
\begin{equation}
    \ket{\psi_k} = P_S\ket{\psi_k}+(I-P_S)\ket{\psi_k} \rightarrow \ket{\psi}.
\end{equation}
Since each $\ket{\psi_k}$ is strictly localized to $J(\mathcal{D})$, we have, from the continuity of the inner product, that
\begin{equation}
    \lim_k\bra{\psi_k}A\ket{\psi_k} = \bra{\psi}A\ket{\psi} = \bra{\Omega}A\ket{\Omega}
\end{equation}
for all $A\in\mathcal{A}(\mathcal{D}')$. Therefore, $\ket{\psi}$ is strictly localized to $J(\mathcal{D})$. However, this contradicts the result that no unit vector in $S$ is strictly localized. We conclude that the inequality in \eqref{eq_positive_gap} holds.

Using the separation between $S$ and $\Psi$, we can show that $\rho$ has a positive trace distance from all states in $\overline{\operatorname{co}(\Psi)}$. Consider any $\sigma \in \operatorname{co}(\Psi)$. Then, we have:
\begin{equation}
    \tr(P_S \sigma ) = \sum_k p_k \bra{\psi_k}P_S\ket{\psi_k} \leq c \sum_k p_k = c.
\end{equation}
However, since $\tr(P_S \rho) = 1$, we get a strictly positive trace distance
\begin{equation}
    D(\rho,\sigma) \geq \abs{\tr(P_S \rho) - \tr(P_S \sigma )} \geq 1 - c > 0.
\end{equation}
From the continuity of the trace, we also have that the inequality holds for all $\sigma\in \overline{\operatorname{co}(\Psi)}$. Hence, $\rho$ has a positive trace distance from every state in $\overline{\operatorname{co}(\Psi)}$ as well.

Trace distance quantifies the optimal distinguishability of two states. The result above, therefore, shows that the separation between $\rho$ and mixtures of strictly localized pure states is not merely formal. In principle, there exists a measurement that distinguishes them with a uniform gap. In the next section, we discuss how such states can arise from more general measurement schemes.

\section{Unsharp Local Measurements\label{sec_5}}

In \cref{sec_3}, we showed that although selective post-measurement states of local, projective measurements are not strictly localized, the corresponding non-selective states are expressible as mixtures of strictly localized pure states. However, in \cref{sec_4}, we found that this does not exhaust the space of strictly localized mixed states, as there exist strictly localized states that cannot be approximated by mixtures of strictly localized pure states. The purpose of this section is to explore how such states can be generated.

To do this, we must consider more general measurement schemes. Any Licht map can be realized by introducing a suitable ancilla \cite{stinespring1955, ozawa1984}. Preparing the ancilla in a fixed state $\ket{0}$, we choose a unitary $U$ satisfying:
\begin{equation}\label{eq_Licht_map_through_ancilla}
    U\left( \ket{\psi} \otimes \ket{0} \right) = \sum_r W_r \ket{\psi} \otimes \ket{r},
\end{equation}
where $\ket{\psi}$ is any state in the original Hilbert space, and $\ket{r}$ are orthonormal ancilla states. The Licht map is recovered by tracing out the ancilla after the unitary evolution.

We want to implement the Licht map introduced in \cref{sec_4}. Its Kraus operators are of the form
\begin{equation}\label{eq_W_r}
    W_r = \sum_k \alpha_{rk}P_k, \quad \sum_r |\alpha_{rk}|^2=1,
\end{equation}
where $\alpha_{rk}$ are coefficients and $P_k$ are mutually orthogonal projectors that sum to the identity. They may, for example, be the spectral projectors of a local observable
\begin{equation}
    A=\sum_k \lambda_kP_k.
\end{equation}
Inserting \eqref{eq_W_r} into \eqref{eq_Licht_map_through_ancilla}, we obtain
\begin{equation}
    U\left(  \ket{\psi} \otimes \ket{0}  \right) = \sum_k P_k\ket{\psi}\otimes\ket{\eta_k}, \quad\ket{\eta_k}=\sum_r\alpha_{rk}\ket r.
\end{equation}
Thus, the probe state $\ket{\eta_k}$ records the outcome $\lambda_k$ of the underlying observable. 

Measuring the probe in the basis $\ket{r}$ induces a positive operator-valued measure (POVM) on the system, whose elements, also called effects, are
\begin{equation}
E_r=W_r^\dagger W_r=\sum_k|\alpha_{rk}|^2P_k.
\end{equation}
Whenever the effects $E_r$ are not projectors, the probe readout constitutes an unsharp measurement of $A$ \cite[Chap. 9.3]{busch2016}.

However, if we ignore the probe, then for the vacuum state, we obtain
\begin{equation}
\begin{aligned}
    \tr_{\mathrm{anc}}\left(U (\Omega \otimes \ket{0}\bra{0}) U^\dagger\right) &= \sum_{k,l}\braket{\eta_l}{\eta_k}P_k\Omega P_l \\
    &= \sum_r W_r\Omega W_r^\dagger.
\end{aligned}
\end{equation}
This is equivalent to measuring the probe in the basis $\ket{r}$ and then forgetting the outcome. The overlap $\braket{\eta_l}{\eta_k}$ determines how the off-diagonal blocks $P_k\Omega P_l$ are modified. Recall that these blocks represent coherence between the possible outcomes $\lambda_k$ and $\lambda_l$ of the underlying observable $A$.

This gives a simple physical interpretation of the example state in \cref{sec_4}. If the probe states $\ket{\eta_k}$ are mutually orthogonal, they provide perfectly distinguishable records of the outcomes of $A$, and tracing out the probe produces complete decoherence in the non-selective state:
\begin{equation}
    \tr_{\mathrm{anc}}\left(U (\Omega \otimes \ket{0}\bra{0}) U^\dagger\right) = \sum_k P_k\Omega P_k.
\end{equation}
We then recover the situation considered in \cref{sec_3}. On the other hand, if the probe states are not mutually orthogonal, they do not provide perfectly distinguishable records, and coherence between different outcomes of $A$ may remain.

The non-selective state constructed in \cref{sec_4} arises precisely from this second situation. In the ancilla representation, the probe becomes entangled with the field, but it does not retain perfectly distinguishable records of the possible outcomes of the local observable. The corresponding probe readout induces an unsharp observable. When the probe is ignored, the field is left in a strictly localized mixed state that cannot be understood as ordinary ignorance over strictly localized pure states.

\section{Discussion and Conclusion\label{sec_6}}

Strictly localized states are physically important for two reasons. First, since they look like vacuum outside the localization region, they can describe states made locally on demand. In that sense, they are consistent with causality. Second, they arise naturally from local operations. In particular, applying a local trace-preserving operation, or more generally a Licht map, to the vacuum produces a strictly localized state. If the map has a single Kraus operator, the resulting state is pure, whereas a general Licht map may produce a strictly localized mixed state.

In several important cases, strictly localized mixed states reduce to mixtures of strictly localized pure states. For example, given a local, projective measurement, even though the selective post-measurement states are not strictly localized, the corresponding non-selective state can be written as a mixture of strictly localized pure states. More generally, whenever the decoherence can be represented as an average over local unitaries, the resulting state is a mixture of strictly localized pure states.

However, this does not exhaust the space of strictly localized mixed states. We have constructed strictly localized mixed states that cannot be expressed, or even approximated in trace distance, by mixtures of strictly localized pure states. Thus, mixtures of strictly localized pure states do not capture the full range of strictly localized mixed states.

The ancilla representation relates these more general states to entanglement with a probe. If tracing out the probe gives complete decoherence, then we recover a mixture of strictly localized pure states. However, if the measurement is unsharp, then discarding the probe can yield a strictly localized mixed state that cannot be understood as classical ignorance over strictly localized pure states.

These results sharpen our intuition about observables, measurements, and locality. In particular, the post-measurement state is determined not by the observable alone, but also by how its measurement is implemented. As shown in \cref{sec_5}, the non-selective state in Eq.~\eqref{eq_rho_finite_discrete} arises from a sharp measurement, whereas the state constructed in \cref{sec_4} arises from an unsharp measurement of possibly the same underlying observable. Nevertheless, their localization properties differ fundamentally. The latter cannot be approximated by mixtures of strictly localized pure states, whereas the former admits exactly such a decomposition.

The appendices contain two supporting results. In \cref{app_1}, we extend Knight's theorem to mixed states by showing that strictly localized bosonic mixed states contain terms with arbitrarily high particle number. In \cref{app_2}, we relate Licht maps to the fixed-point formalism. Specifically, Licht maps associated with some localization region are precisely the trace-preserving operations whose dual leaves every observable outside the localization region fixed.

\section*{Acknowledgments}

The authors used ChatGPT (OpenAI; various GPT-5 models) as an aid in discussing scientific questions, independently checking some arguments and derivations, and improving the presentation of the manuscript. The authors take full responsibility for the content of the article. This work was supported by the Research Council of Norway [grant number 354574].

\appendix
\crefalias{section}{appendix}
\section{Extending Knight's Theorem\label{app_1}}

Knight demonstrated that strictly localized bosonic vector states other than vacuum must contain terms with arbitrarily high particle content; i.e., they cannot have finite particle-number support \cite{knight1961}. Concretely, consider the non-vacuum state
\begin{equation}
    \ket{\psi} = \sum_n c_n\ket{\chi_n}, \quad P_n \ket{\psi} = c_n \ket{\chi_n},
\end{equation}
where $P_n$ is the projector onto the $n$-particle subspace. If there is an $N\in\mathbb{N}$ such that $c_n=0$ for all $n> N$, then $\ket{\psi}$ cannot be strictly localized. In this appendix, we show that a similar conclusion applies to mixed states. To characterize the particle-number support of a potentially mixed state $\rho$, we decompose it in the following way:
\begin{equation}
    \rho = \sum_{n,m}\rho_{nm}, \quad \rho_{nm} = P_n\rho P_m.
\end{equation}
We say that $\rho$ has finite particle-number support if there exists an $M\in \mathbb{N}$ such that $\rho_{nm}=0$ whenever $n > M$ or $m > M$. 

However, note that any vanishing diagonal block forces the corresponding row and column to vanish. Indeed, if $\rho_{NN}=0$, then for any $\ket{\chi_n}$ in the $n$-particle subspace and any $\ket{\chi_N}$ in the $N$-particle subspace, the positivity of $\rho$ and the Cauchy--Schwarz inequality give
\begin{equation}
    \left|\bra{\chi_n}\rho\ket{\chi_N}\right|^2 \leq \bra{\chi_n}\rho\ket{\chi_n} \bra{\chi_N}\rho\ket{\chi_N} = 0.
\end{equation}
Hence, $\rho_{nN}=0$ for all $n$. Similarly, $\rho_{Nm}=0$ for all $m$. Therefore, $\rho$ has finite particle-number support precisely when there is a largest particle number $N$ such that $\rho_{NN}\neq0$, and all diagonal blocks involving higher particle numbers vanish:
\begin{equation}
    \rho_{nn}=0, \quad \forall n>N.
\end{equation}

We also need a formulation of the Reeh--Schlieder property found in \cite{gulla:inprep:a}. Let $f(x)$ be a real smearing function. The corresponding smeared field can be decomposed as
\begin{equation}\label{eq_app_sf}
    \phi_f=\int \dd[4]x \, f(x)\phi(x)=a+a^\dagger,
\end{equation}
where $a$ and $a^\dagger$ are annihilation and creation operators obtained from the usual mode expansion of the field \cite{gulla2023}. The Reeh--Schlieder property can then be formulated in the following way: For any $N$-particle state $\ket{N}$ and region $\mathcal{O}$, there exists a real smearing function $f(x)$ with support in $\mathcal{O}$ such that
\begin{equation}\label{eq_RMP}
    \bra{N}\norder{\phi_f^N}\ket{\Omega} = \bra{N}(a^\dagger)^N\ket{\Omega} \neq 0.
\end{equation}

We are now ready to extend Knight's theorem. Let $\rho$ be strictly localized to some $\mathcal{O}$. Assume, for contradiction, that $\rho$ has nonzero, finite particle-number support. Denote by $N>0$ the largest particle number such that $\rho_{NN}\neq0$ and all blocks involving higher particle numbers vanish. Choose $f(x)$ to be any real test function whose support is outside $\mathcal{O}$, and consider the smeared field in \eqref{eq_app_sf}. From strict localization, we have that
\begin{equation}
    \tr(\rho \norder{\phi_f^{2N}}) = \bra{\Omega}\norder{\phi_f^{2N}}\ket{\Omega} = 0,
\end{equation}
for all such $f(x)$. 

However, we may also evaluate the expectation value of $\norder{\phi_f^{2N}}$ directly. In the normally ordered expansion, the terms have the form $(a^\dagger)^k a^l$, where $k+l=2N$. If $k<N$, then $l>N$, so $a^l$ annihilates every particle sector on which $\rho$ has support. An analogous argument applies to $(a^\dagger)^k$ when $k>N$. Therefore, the only term that can contribute is the term with $k=l=N$, and it acts only on the $\rho_{NN}$ block. Thus,
\begin{equation}\label{eq_app_sl_cond}
    \tr(\rho \norder{\phi_f^{2N}}) = \binom{2N}{N}\tr(\rho_{NN} (a^\dagger)^N a^N) = 0.
\end{equation}
Since $\rho_{NN}$ is positive and has support in the $N$-particle subspace, we may diagonalize it using $N$-particle states $\ket{N_k}$:
\begin{equation}
    \rho_{NN} = \sum_k p_k \ket{N_k}\bra{N_k}, \quad p_k>0.
\end{equation}
Inserting this into \eqref{eq_app_sl_cond}, we obtain
\begin{equation}
\begin{aligned}
    \tr(\rho_{NN} (a^\dagger)^N a^N) &= \sum_k p_k \bra{N_k}(a^\dagger)^N a^N\ket{N_k} \\
    &= \sum_k p_k \norm{a^N\ket{N_k}}^2 = 0.
\end{aligned}
\end{equation}
Since each term is non-negative, this implies that
\begin{equation}
    a^N\ket{N_k} = 0,
\end{equation}
for all $k$ and all real $f(x)$ with support outside $\mathcal{O}$. This contradicts \eqref{eq_RMP}, which implies that there exists a real test function $f(x)$, supported outside $\mathcal{O}$, such that
\begin{equation}
    a^N\ket{N_k}\neq0.
\end{equation}
Therefore, we conclude that mixed strictly localized bosonic states cannot have finite particle-number support.

\section{Licht Maps Through Fixed Points\label{app_2}}

In this section, we show how Licht maps relate to the fixed point formalism. A similar result can also be found in \cite{mandrysch2026}. Several characterizations of fixed points exist in the literature \cite{Luders1951, busch1998, lubnauer2018, arias2002}. For a Hilbert space $H$, define $\mathcal{B}(H)$ and $\mathcal{T}(H)$ to be the sets of bounded and trace-class operators on $H$, respectively. Let $\mathcal{W}: \mathcal{T}(H)\rightarrow\mathcal{T}(H)$ be an operation with Kraus operators $W_k$. We may consider the dual picture (or generalized Heisenberg picture) $\mathcal{W}^*: \mathcal{B}(H) \rightarrow \mathcal{B}(H)$ by
\begin{equation}
    \mathcal{W}^*(A) = \sum_k W_k^\dagger A W_k.
\end{equation}
This map occurs whenever we calculate the expected value of $A\in\mathcal{B}(H)$. That is, given some state $\mathcal{W}(\rho)$, we have
\begin{equation}
    \tr(A\mathcal{W}(\rho)) = \tr(\sum_k W_k^\dagger AW_k \rho) = \tr(\mathcal{W}^*(A) \rho).
\end{equation}
We say that $A$ is a fixed point of $\mathcal{W}^*$ whenever $\mathcal{W}^*(A) = A$.

We know that for a state $\rho$, strictly localized to some $\mathcal{O}$, there exists a Licht map $\mathcal{W}$ such that
\begin{equation}
    \mathcal{W}(\Omega) = \sum_kW_k\Omega W_k^\dagger= \rho,
\end{equation}
where $W_k\in\mathcal{A}(\mathcal{O}^C)'$ and $\sum_k W_k^\dagger W_k = 1$. We notice that the dual map satisfies
\begin{equation}\label{eq_fixed_point_dual_map}
    \mathcal{W}^*(A) = A,  \quad \forall A\in\mathcal{A}(\mathcal{O}^C).
\end{equation}
In other words, every element of $\mathcal{A}(\mathcal{O}^C)$ is fixed by $\mathcal{W}^*$. In this appendix, we will show that the converse is true as well. That is, assume that \eqref{eq_fixed_point_dual_map} holds; then $W_k\in\mathcal{A}(\mathcal{O}^C)'$ for all $k$.

By assumption, each $A\in \mathcal{A}(\mathcal{O}^C)$ is a fixed point. Hence, for a given $A\in\mathcal{A}(\mathcal{O}^C)$, the operators $A^\dagger$ and $A^\dagger A$ are also fixed points. Consider the following operator:
\begin{equation}
    S = \sum_k [A, W_k]^\dagger [A, W_k]. 
\end{equation}
Writing out the expression, we get
\begin{equation}
    S = \mathcal{W}^*(A^\dagger A) - \mathcal{W}^*(A^\dagger)A - A^\dagger \mathcal{W}^*(A) + A^\dagger A = 0.
\end{equation}
Since each term in the sum defining $S$ is positive, $S=0$ implies
\begin{equation}
    [A,W_k]^\dagger[A,W_k]=0, \quad \forall k.
\end{equation}
Hence, $[A,W_k]=0$ for every $k$.

Thus, for a trace-preserving completely positive map $\mathcal{W}$, the following two conditions are equivalent:
\begin{equation} 
\mathcal{W}^*(A)=A, \quad \forall A\in\mathcal{A}(\mathcal{O}^C),
\end{equation}
and
\begin{equation}
W_k\in\mathcal{A}(\mathcal{O}^C)', \quad \forall \text{ Kraus operators }W_k.
\end{equation}
Therefore, Licht maps are precisely the trace-preserving operations whose dual leaves every observable outside the localization region fixed.

\bibliography{refs}
\end{document}